\documentclass[runningheads,a4paper]{llncs}
\usepackage{booktabs}
\usepackage{adjustbox}
\usepackage{threeparttable}
\usepackage{orcidlink}
\usepackage{amssymb}
\usepackage{amsmath}
\usepackage{graphicx}
\usepackage{tikz}
\usepackage{tikz}
\usetikzlibrary{chains, quotes, positioning}
\usepackage{hyperref} 
\hypersetup{
    colorlinks=true, 
    citecolor=blue, 
    linkcolor=black, 
    urlcolor=blue 
}

\usepackage[most]{tcolorbox}
\usepackage[frozencache,cachedir=.]{minted}
\usepackage{enumitem}
\usepackage{subcaption}
\usepackage{diagbox}
\usepackage{tabularx}

\usepackage{url}
\urldef{\mailsa}\path|{alfred.hofmann, ursula.barth, ingrid.haas, frank.holzwarth,|
\urldef{\mailsb}\path|anna.kramer, leonie.kunz, christine.reiss, nicole.sator,|
\urldef{\mailsc}\path|erika.siebert-cole, peter.strasser, lncs}@springer.com|

\begin{document}

\mainmatter  

\title{Vulnerability Detection: From Formal Verification to Large Language Models and Hybrid Approaches: A Comprehensive Overview}

\titlerunning{Vulnerability Detection: From Formal Verification to Large Language Models}

\author{
  Norbert Tihanyi\orcidlink{0000-0002-9002-5935}\inst{1,2} \and
  Tamas Bisztray\orcidlink{0000-0003-2626-3434}\inst{3,4} \and
  Mohamed Amine Ferrag\orcidlink{0000-0002-0632-3172}\inst{5} \and
Bilel Cherif\orcidlink{0009-0006-0095-106X}\inst{2}\and 
Richard A. Dubniczky\orcidlink{0009-0003-3951-1932}\inst{1}\and 
  Ridhi Jain\orcidlink{0000-0002-6102-7114}\inst{2} \and
  Lucas C. Cordeiro\orcidlink{0000-0002-6235-4272}\inst{6,7}
}

\institute{%
\inst{} Eötvös Loránd University (ELTE), Budapest, Hungary \\
\and
\inst{} Technology Innovation Institute (TII), Abu Dhabi, UAE \\
\and
\inst{} University of Oslo, Oslo, Norway \\
\and
\inst{} Cyentific AS, Oslo, Norway \\
\and
\inst{} Guelma University, Guelma, Algeria \\
\and
\inst{} The University of Manchester, Manchester, UK \\
\and
\inst{} Federal University of Amazonas, Manaus, Brazil \\
}
\authorrunning{N.~Tihanyi et al.
}

\toctitle{Vulnerability Detection with LLMs and FV tools}
\tocauthor{N.Tihanyi et al.}
\maketitle

\begin{abstract}
Software testing and verification are critical for ensuring the reliability and security of modern software systems. Traditionally, formal verification techniques, such as model checking and theorem proving, have provided rigorous frameworks for detecting bugs and vulnerabilities. However, these methods often face scalability challenges when applied to complex, real-world programs. Recently, the advent of Large Language Models (LLMs) has introduced a new paradigm for software analysis, leveraging their ability to understand insecure coding practices. Although LLMs demonstrate promising capabilities in tasks such as bug prediction and invariant generation, they lack the formal guarantees of classical methods. This paper presents a comprehensive study of state-of-the-art software testing and verification, focusing on three key approaches: \textit{classical formal methods}, \textit{LLM-based analysis}, and \textit{emerging hybrid techniques}, which combine their strengths. We explore each approach's strengths, limitations, and practical applications, highlighting the potential of hybrid systems to address the weaknesses of standalone methods. We analyze whether integrating formal rigor with LLM-driven insights can enhance the effectiveness and scalability of software verification, exploring their viability as a pathway toward more robust and adaptive testing frameworks.
\end{abstract}

\section{Introduction}

\begin{figure}[t]
    \centering
    \includegraphics[width=\linewidth]{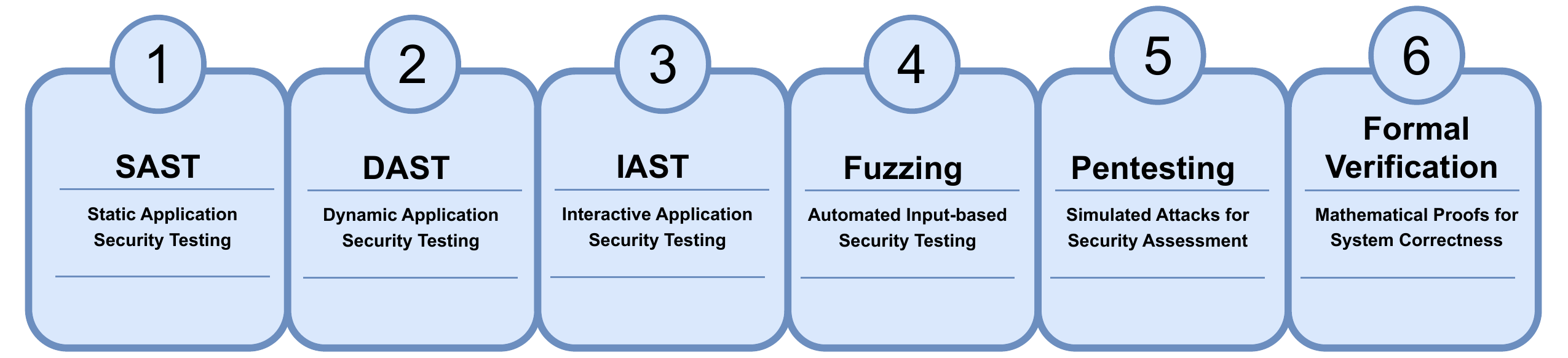}
    \caption{High-level categories of Software Testing}
    \label{fig:Maincategories}
\end{figure}

Software testing is critical to software development, ensuring reliability, security, and performance. Various testing techniques are employed to detect bugs, vulnerabilities, and inefficiencies throughout different development stages. In the literature, software testing is categorized into different groups. Figure~\ref{fig:Maincategories} illustrates the six primary high-level categories of software testing. \textit{Static Application Security Testing (SAST)} inspects code without executing it, detecting potential issues early in development~\cite{staticanalysis2022}. \textit{Dynamic Application
Security Testing (DAST)} validates software behavior during runtime, identifying vulnerabilities that may only surface under real-world conditions. \textit{Interactive Application
Security Testing (IAST)} blends insights from both static and dynamic analysis by instrumenting the application to observe security issues during execution~\cite{IAST}. \textit{Fuzzing} leverages unexpected or malformed inputs to uncover hidden weaknesses~\cite{fuzzing}. In contrast, \textit{Penetration Testing} (often a manual or semi-automated process) simulates attacker behaviors to find exploitable security flaws in realistic scenarios~\cite{pentest}. \textit{Formal verification (FV)} is a rigorous mathematical approach to proving that hardware or software systems fulfill their intended requirements under all possible scenarios~\cite{Formalverification}. Unlike traditional testing that only covers some executions, FV either formally proves a system meets its properties or provides a concrete counterexample, eliminating false positives.

These six categories provide a broad perspective on modern application security strategies. However, some sources may classify them differently, such as grouping fuzzing under dynamic testing or recognizing only three primary categories—SAST, DAST, and IAST—while others introduce new categories like \textit{Run-time Application Security Protection (RASP)}~\cite{RASP}. However, this review will treat these six as the main categories, since each demands specialized expertise and distinct methodologies that naturally separate them based on the required knowledge.

Lately, \textit{Large Language Models (LLMs)} have been introduced and are now used in every part of software testing, from fuzzing~\cite{LLMfuzz} to invariant generation~\cite{invariants}, and also play an important role in the generation of source code. As LLMs are expected to dominate future software development, the need for robust automated testing has become more urgent, as multiple studies reveal that LLMs can introduce security vulnerabilities in all programming languages examined~\cite{tihanyi_how_2024},~\cite{wang2024your},~\cite{roychoudhury2025ai},~\cite{lyu2024automatic}. In the field of software testing, FV and LLMs present a promising combination for addressing these challenges. Each offers unique strengths and weaknesses, but they can complement and compensate for one another’s limitations.

Among the various FV techniques, \emph{Bounded Model Checking} (BMC)~\cite{BMCfirst} has gained particular prominence for its scalability. Rather than exhaustively analyzing the entire state space, BMC confines its search to a predefined bound, employing powerful satisfiability (SAT/SMT) solvers to detect property violations. 
While this approach excels at uncovering memory safety and concurrency issues, it fails to address various vulnerabilities, such as SQL injection or cross-site scripting (XSS).
Moreover, the \emph{state explosion problem} can make verification computationally infeasible for large or highly concurrent systems. Finally, most FV tools focus on a small set of languages---primarily C, C++, and Java, with python being recently introduced~\cite{10.1145/3650212.3685304}---limiting their relevance for modern software stacks. 

In parallel, LLMs have gained significant traction in software security research. These models use neural architectures trained on massive source code and natural language corpora to detect vulnerabilities. Unlike traditional static or dynamic analysis, which relies on handcrafted rules or symbolic execution, LLM-based approaches leverage patterns that generalize across different programming languages and codebases. However, current LLMs face challenges: they may overfit training data, generate false positives, and lack the rigorous guarantees formal methods offer. Nonetheless, as LLMs mature and incorporate advanced learning paradigms, they promise to complement FV techniques by providing broader coverage of language-specific and domain-specific vulnerabilities. This review article offers the following key contributions:
\begin{enumerate}
\small
    \item \textbf{Analysis of Formal Verification--Based Methods.} We provide a comprehensive overview of how FV techniques are used to detect vulnerabilities, outlining their current capabilities and the challenges---such as scalability and limited language support---that constrain their broader adoption.

    \item \textbf{Assessment of LLM-Based Approaches.} We critically examine cutting-edge research leveraging large language models for automated vulnerability detection, highlighting the advantages of learned representations and discussing ongoing challenges such as false positives and domain adaptation.

    \item \textbf{Exploration of Hybrid Strategies.} We investigate emerging frameworks that integrate FV with LLM-based methods, illustrating how both strengths can be combined to achieve broader coverage, higher accuracy, and greater scalability in vulnerability detection.
\end{enumerate}

\section{Motivation}

\begin{figure}[ht]
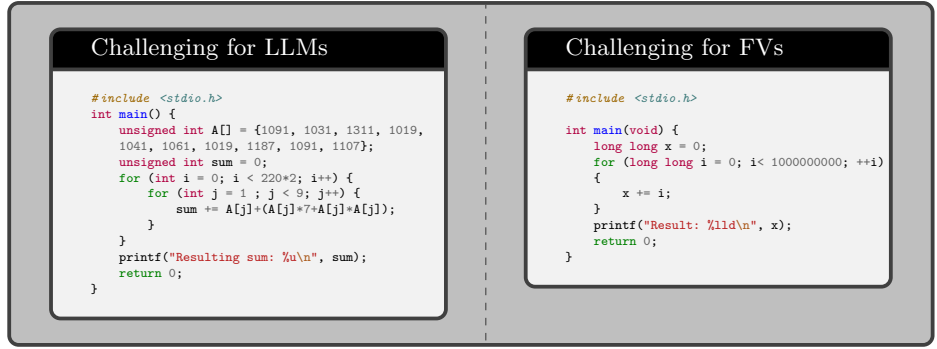
\centering
\begin{tcolorbox}[sidebyside, 
                  lefthand width=.43\textwidth,
                  sidebyside align=top seam,
                  colback=lightgray]

\begin{tcolorbox}[title={Challenging for LLMs }, coltitle=white, colbacktitle=black, colback=gray!10]
\tiny
\begin{minted}[escapeinside=||]{c}
#include <stdio.h>
int main() {
    unsigned int A[] = {1091, 1031, 1311, 1019, 
    1041, 1061, 1019, 1187, 1091, 1107};
    unsigned int sum = 0;
    for (int i = 0; i < 220*2; i++) {
        for (int j = 1 ; j < 9; j++) { 
            sum += A[j]+(A[j]*7+A[j]*A[j]);
        }
    }
    printf("Resulting sum: %u\n", sum);
    return 0;
}
\end{minted}
\end{tcolorbox}

\tcblower

\begin{tcolorbox}[title={Challenging for FVs }, coltitle=white, colbacktitle=black, colback=gray!10]
\tiny

\begin{minted}[escapeinside=||]{c}
#include <stdio.h>

int main(void) {
    long long x = 0;
    for (long long i = 0; i< 1000000000; ++i)
    {
        x += i;     
    }
    printf("Result: %lld\n", x);
    return 0;
}

\end{minted}
\end{tcolorbox}

\end{tcolorbox}
\caption{Motivating examples posing challenges for LLMs an FV tools }
\label{fig:motivation}

\end{figure}

While LLMs can tackle increasingly complex tasks, they rely on pattern recognition rather than true semantic understanding. Although this makes them versatile as static code analyzers, their outputs carry no formal correctness guarantees. ``Hallucinations'' or plausible looking yet incorrect answers, remain a well-documented problem. Tihanyi et al.~\cite{10825051} observed that when presented with a seemingly simple question—such as finding the next prime number, which would typically require tool usage—an LLM attempted to solve it directly rather than skipping it, ultimately producing an incorrect response. This example illustrates how LLMs can misinterpret problem complexity and face difficulties with exact calculations when the necessary tools are unavailable. However, LLMs can sometimes leverage pattern recognition to determine that exhaustive checks are unnecessary, allowing them to arrive quickly at a correct solution in cases where FV tools might struggle. To illustrate both aspects, we present two small C programs in Figure~\ref{fig:motivation}. The code on the left poses challenges for LLMs, while the one on the right presents difficulties for FV tools.

We tested the latest chat models, including OpenAI ChatGPT-4o, the o1 model, and Deepsek R1’s chat model, with Deepthink enabled. We used the following prompt: ``\textit{You are a security analyst. I want you to analyze the following code for vulnerabilities. For the vulnerabilities found, return the vulnerable line(s) and the corresponding CWE identifier(s). If none is found, return an empty list.''}. 

In every case, for the code on the left-hand side, all models detected a sum overflow and reported CWE-190 (Integer Overflow or Wraparound). However, this is incorrect: the sum comes close to overflowing but does not exceed the maximum limit of \(2^{32}\). Verification is straightforward using an FV tool. In particular, we employed the \textit{Efficient SMT-based Context-Bounded Model Checker (ESBMC)}~\cite{ESBMC}, which uses symbolic model checking based on Satisfiability Modulo Theories (SMT) in combination with abstract interpretation techniques. After just 0.73 seconds on a regular laptop, ESBMC reported \texttt{VERIFICATION SUCCESSFUL}, indicating no overflow was detected. On the other hand, formal verification tools, including ESBMC have limitations. When examining the other program in Figure~\ref{fig:motivation} (right-side), it is straightforward for an LLM to recognize the absence of vulnerabilities in the code, whereas ESBMC requires significantly more time to complete its verification. It needs to unroll the loop $10^9$ times. In our test environment, unwinding the first $1$ million iterations took 46 seconds, implying that fully unrolling the loop would take approximately 12.7 hours to verify, with extremely high memory requirements. These examples underscore that LLMs and formal verification each play a vital role in vulnerability detection. In the following chapters, we explore the latest advances in LLM-based analysis and formal verification methods, illustrating how these tools can be used together to enhance reliability, speed, and accuracy.

\section{LLMs in Vulnerability Detection}

Recent developments, particularly in decoder-only models like OpenAI’s ChatGPT or Meta’s Code Llama, highlight the shifting landscape in vulnerability detection. Their capacity for large-context reasoning and on-demand text generation facilitates powerful few-shot or prompt-based approaches that can outperform classical fine-tuned detectors on certain benchmarks. This section reviews the latest trends in LLM-based vulnerability detection.
\subsection{Model Architectures and Trends}

Vulnerability detectors leverage transformer-based code models as foundational backbones trained on large code corpora across multiple programming languages. While it allows them to learn syntax, the training data itself is full of vulnerable code in many instances. Transformer-based code models can be put in three categories: Encoder-Only, Encoder-Decoder and Decoder-Only. Encoder-only and encoder–decoder transformers (e.g., CodeBERT, GraphCodeBERT, CodeT5) were widely used in early work, often fine-tuned for classification tasks. However, decoder-only LLMs like the GPT series and CodeLlama have rapidly become the preferred choice due to their superior size and generalization capabilities~\cite{sheng2025large}. 

This trend toward decoder-only architectures aligns with developments in industry, where state-of-the-art LLMs are often deployed via in-context learning and prompt engineering rather than full retraining. Large models like GPT-4 can usually be prompted to detect vulnerabilities without any parameter updates, thanks to knowledge gained during pre-training. Advances in prompt design (e.g., chain-of-thought prompts or specialized code analysis prompts) further boost their performance, allowing these models to reason about code security with minimal task-specific tuning~\cite{MECHRI2025104151}. Encoder-only models (e.g. BERT-derived code models) typically require full fine-tuning to perform vulnerability classification, as they lack the broad intrinsic knowledge and generative prowess of frontier LLMs.
The emerging consensus is that while encoders excel at efficient classification, decoder-only LLMs offer a more flexible foundation for vulnerability detection – able to understand complex code patterns and suggest fixes – making them increasingly central in recent research.

\subsection{Datasets for Fine-Tuning and Benchmarking}

LLMs pre-trained on code have been fine-tuned for vulnerability detection with notable success.  Large-scale datasets are mostly available in C, C++, Java, Python, PHP, or Solidity, with C/C++ being the overwhelming majority~\cite{sheng2025large}. Labeled datasets with over $100$ samples are presented in Table~\ref{tab:datasets}. Such datasets are essential for fine-tuning and benchmarking various static, dynamic, or AI/ML-based detection systems. ReGVD combined CodeBERT with graph neural networks to learn from code property graphs, improving the detection of vulnerabilities in C/C++ code~\cite{jiang2024investigating}.
A RoBERTa-based model (VulBERTa) fine-tuned on multiple datasets achieved top-tier results on benchmarks like CodeXGLUE’s defect detection~\cite{hanif2022vulberta}.
A notable result is that SecureFalcon, an innovative model architecture with only 121 million parameters derived from the Falcon-40B model and explicitly tailored for classifying software vulnerabilities, outperforms existing models such as BERT, RoBERTa, CodeBERT, and traditional ML algorithms~\cite{ferrag2023securefalcon}. 
These studies indicate that customizing LLMs using real-world code vulnerability datasets significantly enhances their detection effectiveness in practical applications~\cite{du2024generalization}. Overall, fine-tuning remains the dominant approach (around 73\% of recent studies) for adapting LLMs to this task~\cite{zhou2024large}.

\begin{table}[t]
\scriptsize
\centering
\addtolength{\tabcolsep}{-2.3pt}
\caption{Datasets for Benchmarking and Fine-Tuning in Vulnerability detection}
\begin{tabular}{>{\ttfamily}p{2.7cm} >{\ttfamily}p{1cm} >{\ttfamily}p{2,6cm} >{\ttfamily}p{2.3cm} >{\ttfamily}p{2cm} >{\ttfamily}p{2cm}}
\toprule
\textbf{Dataset} & \textbf{Size*} & \textbf{Language} & \textbf{\#Unique Vuln.} & \textbf{Granularity} & \textbf{Source} \\
\midrule
Draper$^*$ \cite{russell2018automated} & 1274k & C/C++& 5 & function & mixed \\
SARD \cite{NIST_SARD} & 450k & C/C++,Java, PHP & 150 & file & mixed \\
SeVc \cite{li2022sevc} & 420k & C/C++ & 126 & function & mixed \\
DiverseVul \cite{chen2023diversevul} & 349k& C/C++ & 150 & function & real-world \\
FormAIv2 \cite{tihanyi_how_2024} & 331k& C & 33 & file & AI Generated \\
BigVul \cite{fan2020bigvul} & 264k  & C/C++ & 91 & function & real-world \\
PrimeVul \cite{ding_vulnerability_2024} & 235k  & C/C++ & 140 & function & real-world \\
VulDeeLocator \cite{li2022_vuldeelocator} & 198k  & LLVM IR & 4 & function & mixed \\
Juliet C/C++ \cite{2024nistjulietc} & 64k & C/C++ & 118 & file & by design vuln. \\
Code Gadgets \cite{li2018vuldeepecker} & 61k& C/C++ & 2 & function & mixed \\
Juliet Java \cite{2024nistjulietjava} & 28k & Java & 112 & file & by design vuln. \\
FELLMVP \cite{luo2024fellmvp} & 15k& Solidity & 8 & contract & real-world \\
CVEfixes \cite{guru2021cvefixes} & 12,k & C/C++,Python, PHP & 180 & commit & real-world \\
ChatPHP\cite{10.1007/978-3-031-68738-9_34} & 2,5k & PHP & 4 & website & AI Generated \\
VCMatch \cite{wang2022_vcmatch} & 1,6k & C/C++,Java,PHP & 7 & commit & real-world \\
SVEN \cite{he2023_large} & 1,6k & C/C++,Python & 9 & commit & real-world \\
Ponta2019 \cite{ponta2019_manually} & 1,2k  & Java & 6 & commit & real-world \\
VulBench \cite{gao_how_2023} & 455 & C/C++& 9 & function & mixed \\
Ullah2023 \cite{ullah_llms_2024} & 228 & C/C++, Python & 8 & function & mixed \\
Smartbugs\cite{ferreira2024-smartbugs} & 143 & Solidity & 10 & contract & mixed \\
MAGMA \cite{ahmad2024_magma} & 138 & C/C++, Lua, PHP & 11 & repository & real-world \\
Vulcorpus \cite{kouliaridis_assessing_2024} & 100  & Java & 10 & function & by design vuln.\\
\bottomrule
\end{tabular}
Legend:  *Draper has been labelled with static analysers and are therefore less reliable.
\footnotesize
\centering
\label{tab:datasets}
\end{table}

\subsection{Prompt-Based Approaches}

With the advent of powerful decoder-only models like GPT-3.5/4, researchers explored prompt-based methods to detect vulnerabilities without gradient-based training. These approaches supply code to a frozen LLM and ask it (via a prompt) whether the code is vulnerable, possibly with explanations. A key finding is that prompt design and reasoning techniques dramatically affect performance~\cite{wei2022chain}. Few-shot or instructed large models can leverage their broad knowledge to detect many vulnerabilities when asked the right way. Prompt-based methods offer quick iteration (no training needed) and flexibility across languages, but they have trade-offs. They often require large context windows to input code and may struggle with very long or complex code snippets due to context length limits. In addition, responses can be inconsistent – one study found that even advanced code models achieved only around 54.5\% balanced accuracy when directly queried, and simply scaling up model size or naive prompting did not significantly improve this.

\subsection{Emerging Trends and Ongoing Challenges}

Despite recent progress, significant challenges remain in using LLMs for vulnerability detection. While fine-tuned models may excel on partially trained datasets, they often fail to generalize to new, unseen data. Decoder-only models, however, have shown strong performance on code they have not encountered before. A major bottleneck for both approaches is data availability: accurately labeled large datasets with diverse vulnerability representations—ones that truly mirror real production environments—are still missing for most programming languages. Moreover, existing datasets tend to be skewed toward certain vulnerability types (e.g., buffer overflows, injections), leaving others (such as concurrency flaws) under-represented~\cite{sheng2025large}. Efforts to crowdsource or generate more diverse samples (including synthetic code produced by LLMs) are underway to enrich training data~\cite{tihanyi_how_2024}.

Another core issue is semantic understanding. Current LLMs often struggle with intricate code logic, such as pointer aliasing or subtle data flows, especially when the vulnerability stems from a long sequence of operations. Studies have shown that small, behavior-preserving changes (e.g., renaming variables or rearranging code) can mislead an LLM’s judgment~\cite{steenhoek2024err,299557}. This raises concerns about reliability and false positives/negatives: a model might confidently mislabel safe code as vulnerable or vice versa. These errors are difficult to debug without a transparent explanation or consistent reasoning. Ensemble or consensus methods can mitigate some of this uncertainty but add system complexity. Scalability and integration pose further hurdles: Large models like GPT-4 are powerful but expensive to run and not easily deployed on-premise. To address this, companies are exploring distilled or domain-adapted versions that run faster and use parameter-efficient fine-tuning methods (e.g., LoRA adapters) to customize massive models for vulnerability detection without fully retraining them.

On the positive side, practical applications of LLM-based vulnerability detection are emerging. Integrated Development Environment (IDE) plugins and code review bots powered by models like Codex can flag risky code in real time and suggest fixes. Security companies have begun building AI “copilots” for auditors—for instance, using an LLM to scan code repositories and highlight high-risk functions for manual review, thereby combining the speed of automation with human judgment. In specialized settings such as blockchain, tools are appearing that apply LLMs to audit smart contracts for pitfalls like reentrance and integer overflow. These developments underscore an evolving human–AI partnership model: the LLM handles large-scale code analysis and reasoning, while human experts validate the results and make final decisions, guided by the AI’s insights.

\section{Formal Verification}

Formal verification (FV) is a rigorous mathematical approach for establishing that hardware and software systems satisfy their intended requirements across \emph{all} possible scenarios. In contrast to traditional testing, which examines only a limited subset of executions, FV can provide either (i) a formal proof that the system meets the specified properties or (ii) a concrete counterexample demonstrating how the properties can be violated. This dual capability ensures that any detected vulnerabilities are backed by a demonstrable failing execution, thereby minimizing false positives.

A widely used technique within FV is \emph{model checking} (MC), introduced independently by Edmund Clarke and Allen Emerson in 1981~\cite{MC1}, and by Jean-Pierre Queille and Joseph Sifakis in 1982~\cite{MC2}. MC systematically explores the state space of a system to verify whether certain properties, typically expressed in temporal logic, hold. It has been successfully applied to various domains, including hardware circuits, concurrent systems, and communication protocols. However, traditional MC often encounters the well-known \textit{state explosion problem}~\cite{Clarke2012}, where the number of states grows exponentially with the increase in variables and concurrency.

To mitigate this complexity, \textit{Bounded Model Checking} (BMC) was introduced in 1999 by Armin Biere, Alessandro Cimatti, Edmund Clarke, and Yunshan Zhu~\cite{BMCfirst} as a complementary approach to Binary Decision Diagrams (BDDs). Rather than exhaustively exploring all possible states, BMC restricts its search to executions up to a predefined bound. This makes BMC more scalable in practice and highly effective at detecting property violations in complex systems, as any counterexample within the bound will be discovered. While BMC can become computationally demanding for massive or deeply nested systems, it remains an invaluable tool for identifying security and reliability issues. For instance, ESBMC was recently employed to analyze the Arm Confidential Computing Architecture, uncovering specification failures that strengthened its security\cite{ARM}. Likewise, BMC techniques have revealed common bugs (e.g., null pointer dereferences, buffer overflows) in large open-source C projects, demonstrating the practical benefits of FV for real-world software~\cite{sousa2023lsverifier}.

An important takeaway for the reader is that these techniques provide formal guarantees: if a bug is identified, a counterexample confirms it, thereby minimizing false positives. This is especially valuable in mitigating the weaknesses of LLMs.

\section{Hybrid Approaches}

Integrating LLMs with formal methods and program analysis tools has emerged as a promising direction for improving software vulnerability detection. Traditional formal verification methods provide rigorous guarantees about software correctness but often struggle with scalability and applicability to large-scale, real-world software systems. LLMs, on the other hand, can rapidly analyze vast amounts of code and detect patterns indicative of vulnerabilities, but lack formal guarantees of correctness and often produce false positives.

\subsection{LLM-Assisted Formal Verification}

LLMs have been applied to assist in writing the formal artifacts (properties, invariants, constraints) needed for verification. For example, Kande et al. (2023) investigate using GPT-based models to generate security assertions for hardware designs~\cite{10458667}. Similarly, Pei et al. (ICML 2023) showed that LLMs fine-tuned for program invariant generation can infer loop invariants for code with a quality comparable to dynamic invariant detectors~\cite{pmlr-v202-pei23a}.

A notable application is PropertyGPT by Liu et al. (NDSS 2024), which leverages GPT-4 to automatically generate formal properties (invariants, pre- / post-conditions, temporal rules) for smart contract verification~\cite{liu2024propertygpt}. In experiments on real Ethereum contracts, PropertyGPT achieved 80\% recall of ground-truth properties and successfully detected numerous known vulnerabilities and even 12 zero-day flaws that earned bug bounties. These results illustrate that LLMs, when guided properly, can generate high-quality formal specifications to drive vulnerability discovery.

LLMs can serve as “specification generators” or assistants in formal verification workflows~\cite{lin2025fvel}. By translating natural-language requirements or learned patterns into formal assertions/invariants, they reduce the manual burden and open formal methods to broader use in security. The LLM outputs are not guaranteed correct on their own, but coupling them with formal checks (compilation, model checking, theorem proving) ensures that only valid, useful properties are utilized.

Overall, LLM-assisted formal verification has shown success in domains ranging from hardware security to smart contracts, suggesting a general paradigm in which LLM proposes, and formal tools dispose~\cite{liu2024propertygpt}. Another line of work makes the LLM itself part of a verification loop, ensuring its vulnerability analyses adhere to formal correctness criteria. In these approaches, formal verification tools guide or validate the LLM’s reasoning in real-time. CryptoFormalEval is a representative example focusing on cryptographic protocol security~\cite{curaba2024cryptoformaleval}. 

Formal feedback helps the LLM adjust its strategy; for instance, if the prover finds a counterexample, the LLM knows a real vulnerability exists and can elaborate on it; if not, the LLM refines its queries. This tight interaction between LLM and prover means any vulnerability the LLM ``finds” is backed by a formal proof of concept, aligning the outcome with sound verification principles. Beyond specific applications, researchers are formalizing how an LLM can operate under the guidance of a verifier. Wu et al. (2023) propose LEMUR, a framework that interleaves LLM-generated reasoning steps with checks by automated solvers~\cite{wu2023lemur}. They treat the LLM as an abstract transition relation on program states: the LLM suggests a candidate invariant or lemma, then an SMT solver or model checker verifies if it holds. If it passes the formal check, the system accepts the LLM’s suggestion as part of a proof; otherwise, the LLM is prompted to refine its output. Crucially, LEMUR includes a formal calculus with derivation rules for this process and proves it is sound (i.e., it will never accept an incorrect property if the underlying verifier is sound). This formal verification-guided approach ensures the LLM’s creative reasoning is always tethered to correctness. It was shown to improve automated verification on benchmark programs, solving tasks that neither the LLM nor the verifier could easily achieve alone. As this area evolves, we expect LLMs to become increasingly constrained by specs or formal rules during generation (for example, using model checking results as feedback in the LLM’s prompt), thereby producing analyses that are not only plausible but also provably correct.

\subsection{End-to-End Hybrid Pipelines}

End-to-end systems have been built that seamlessly chain together LLM components with traditional analysis tools to detect (and even fix) vulnerabilities. These pipelines treat LLMs and formal/static analyzers as complementary stages, often in an iterative loop. ESBMC-AI~\cite{tihanyi2023new} is a notable framework that illustrates this integration: a self-healing software pipeline that combines a BMC tool with an LLM to detect and repair vulnerabilities in C code. This closed-loop approach was evaluated on 50,000 C programs containing vulnerabilities (from the FormAI dataset). ESBMC-AI was able to automatically detect and repair issues like buffer overflows, arithmetic errors, and null-pointer dereferences with high accuracy. The authors report a significant portion of the bugs were fixed on the first attempt by the LLM, and the remainder eventually resolved through iterative prompting. The ESBMC-AI pipeline demonstrates a powerful synergy: formal analysis provides certainty that a vulnerability exists (and supplies details), while the LLM contributes generative intelligence to suggest a fix, and formal analysis guarantees the fix’s correctness. 

PropertyGPT pipeline~\cite{liu2024propertygpt}, as described earlier, is a system for smart contracts that can also be viewed as an end-to-end pipeline combining retrieval, LLM generation, static checking, and formal verification. The aforementioned CryptoFormalEval~\cite{curaba2024cryptoformaleval}, which can be seen as an end-to-end pipeline in the context of protocol analysis. It provides a middleware for an AI agent (LLM) to interact with a theorem prover in multiple rounds. 

End-to-end hybrid pipelines have shown that combining detection and validation in one loop greatly improves confidence in the results. By cross-checking an LLM’s output with a rigorous tool (and vice versa), they mitigate each other’s weaknesses. For instance, the LLM might introduce a bad fix or a bogus property, but the formal stage will catch it; conversely, a formal tool might miss a bug due to limited specifications, but the LLM can provide those missing pieces earlier in the pipeline. These holistic systems are pioneering initiatives that indicate how future vulnerability detection might be fully automated: an AI finds a bug, explains it, fixes it, and proves the fix correct, all in an integrated workflow~\cite{tihanyi2023new}.

In conclusion, integrating LLMs and FV techniques presents a promising path toward more robust vulnerability detection and software assurance. While FV offers mathematical rigor and precise guarantees, it struggles with scalability and practical applicability to complex systems. On the other hand, LLMs excel in pattern recognition and flexible code analysis but lack formal correctness guarantees. Combining these approaches allows hybrid frameworks to leverage their strengths, using LLMs for broad code analysis and suggestion generation while ensuring correctness through FV validation. Recent research has demonstrated the viability of such hybrid pipelines, paving the way for more efficient, accurate, and scalable software verification methodologies.

\bibliographystyle{plain}  
\bibliography{main} 

\end{document}